\documentclass[11pt,a4paper]{aa}
\usepackage[utf8]{inputenc}
\usepackage[T1]{fontenc}
\usepackage{lmodern}
\usepackage{microtype}
\usepackage{setspace}
\usepackage{graphicx}
\usepackage{booktabs}
\usepackage{hyperref}
\hypersetup{colorlinks=true,linkcolor=blue,citecolor=blue,urlcolor=blue}
\usepackage{titling}
\usepackage{fancyhdr}
\usepackage{amsmath,amssymb}

\usepackage{xcolor}
\usepackage{soul}

\begin{document}

\begin{titlepage}
  \centering
  {\Huge\bfseries Expanding Horizons \\[6pt] \Large Transforming Astronomy in the 2040s \par}
  \vspace{1cm}

  {\LARGE \textbf{The Future of Evolved Planetary Systems}\par}
  \vspace{1cm}

  \begin{tabular}{p{4.5cm}p{10cm}}
    \textbf{Scientific Categories:} & (exoplanets: composition; white dwarfs; stars:  atmospheres, abundances; time-domain astronomy; Galactic archaeology) \\
    \\
    \textbf{Submitting Author:} & Roberto Raddi \\
    & Departament de Física, Universitat Politècnica de Catalunya, c/Esteve Terrades 5, 08860, Castelldefels, Spain  \\
    & roberto.raddi@upc.edu\\
    \\
    \textbf{Contributing authors:} & 
Anna F. Pala$^{1}$,
Alberto Rebassa-Mansergas$^{2,3}$, 
Boris T. G\"ansicke$^4$, 
Lientur Celedon$^{5}$,
Tim Cunningham$^{6}$, 
Camila Damia Rinc\'on$^2$,
Aina Ferrer i Burjachs$^2$,
Enrique Garc\'ia-Zamora$^2$,
Nicola Pietro Gentile Fusillo$^7$, 
Joaquim Meza$^{5}$,
Evelyn Puebla$^{5}$,
Pablo Rodríguez-Gil$^{8,9}$, 
Snehalata Sahu$^4$, 
Alejandro Santos-Garc\'ia$^2$
Odette Toloza$^{5}$, 
Santiago Torres$^{2,3}$, 
Pier-Emmanuel Tremblay$^4$,
Jan van Roestel$^{10}$,
Murat Uzundag$^{11}$,
Dimitri Veras$^{4,12,13}$,
Jamie Williams$^4$\\
\\

\multicolumn{2}{l}{\small $^{1}$European Southern Observatory, Karl Schwarzschild Straße 2, D-85748, Garching, Germany}\\
\multicolumn{2}{l}{\small $^{2}$Departament de Física, Universitat Politècnica de Catalunya, c/Esteve Terrades 5, 08860, Castelldefels, Spain}\\
\multicolumn{2}{l}{\small $^{3}$Institut d'Estudis Espacials de Catalunya (IEEC), C/Esteve Terradas, 1, Edifici RDIT, 08860, Castelldefels, Spain}\\
\multicolumn{2}{l}{\small $^{4}$Department of Physics, University of Warwick, Coventry, CV4 7AL, UK}\\
\multicolumn{2}{l}{\small $^{5}$Departamento de Física, Universidad Técnica Federico Santa María, Avenida España 1680, Valparaíso, Chile}\\
\multicolumn{2}{l}{\small $^{6}$Center for Astrophysics, Harvard \& Smithsonian, 60 Garden St., Cambridge, MA 02138, USA} \\
\multicolumn{2}{l}{\small $^{7}$Universita degli studi di Trieste, Via Valerio, 2, Trieste, 34127, Italy} \\
\multicolumn{2}{l}{\small $^{8}$Instituto de Astrofísica de Canarias, E-38205 La Laguna, Tenerife, Spain}\\
\multicolumn{2}{l}{\small $^{9}$Departamento de Astrofísica, Universidad de La Laguna, E-38206 La Laguna, Tenerife, Spain}\\
\multicolumn{2}{l}{\small $^{10}$Institute of Science and Technology Austria, Am Campus 1, 3400 Klosterneuburg, Austria}\\
\multicolumn{2}{l}{\small $^{11}$Institute of Astronomy, KU Leuven, Celestijnenlaan 200D, Leuven, 3001, Belgium}\\
\multicolumn{2}{l}{\small $^{12}$Centre for Exoplanets and Habitability, University of Warwick, Coventry CV4 7AL, UK}\\
\multicolumn{2}{l}{\small $^{13}$Centre for Space Domain Awareness, University of Warwick, Coventry CV4 7AL, UK}\\

  \end{tabular}

\vspace{1cm}
\begin{minipage}{0.9\textwidth}
\begin{center} \textbf{ABSTRACT} \end{center}

 Understanding the formation, evolution, and chemical diversity of exoplanets are now central areas of astrophysics research.  White dwarfs provide a uniquely sensitive laboratory for studying the end stages of planetary-system evolution and for probing the bulk composition of both rocky and volatile-rich exoplanetary material. In the 2030s new facilities will transform our ability to carry out \textit{``industrial-scale''} astrophysics, leading to fundamental results and new challenges for the next decade. By combining the volume of data surveyed by the ESA {\em Gaia} mission and Vera C. Rubin Observatory with the next-generation of spectroscopic facilities, the European Southern Observatory (ESO) community will be in a position to obtain an unbiased census of evolved planetary systems, constrain the composition of thousands of disrupted planetesimals, and connect these signatures to Galactic populations and stellar birth environments. Thus, it is now the time for assessing those challenges and preparing for the future. This white paper outlines key science opportunities arising in the next decade and the technological requirements of future ESO facilities in enabling transformative discoveries in the 2040s. These future facilities will have to combine a number of features that are crucial for studying evolved planetary systems at white dwarfs, such as broad optical to near-infrared coverage, a high sensitivity at blue wavelengths, multi-resolution capability, massive multi-plexing, and time-domain reactivity.

\end{minipage}

\end{titlepage}


\section{Introduction and Background}
\label{sec:intro}

Over the past twenty years, it has become clear that main-sequence stars commonly host planetary systems \citep{cassan2012}, and that the physical properties of exoplanets and the birth environment of their host stars are closely related \citep{teske2024}. 
While photometric transits and radial-velocity measurements provide constraints on densities \citep{dorn2015} and transmission spectroscopy can measure the atmospheric properties of exoplanets \citep{kreidberg2014}, they do not yield bulk compositions.

White dwarfs, the evolved remnants of Sun-like stars, probe the stellar and planetary evolution after the giant phases \citep{veras2024} and provide direct constraints on the bulk composition of accreted planetary debris \citep{xu2025}. A recent census listed $\approx 1800$ white dwarfs with signs of debris accretion \citep[][]{williams2024}, which is of the same order of magnitude of known exoplanets\footnote{The NASA Exoplanet archive listed 6061 exoplanets on 2025 December 12; \url{https://exoplanetarchive.ipac.caltech.edu/}}. 
\vspace{-0.1cm}
\subsection{The fate of planetary systems} 
After post-main-sequence mass loss, planetary orbits expand and destabilize, scattering the surviving minor bodies and planets that may eventually be tidally disrupted by the white dwarf \citep[][]{jura2008, debes2012,mustill2014,veras2014}. Their debris form long-lived, dusty \citep{jura2003,girven2012, farihi2025} and, in some cases, gaseous discs that trace ongoing accretion onto white dwarfs \citep{gaensicke2006,gaensicke2008}. These structures, analogous to Saturn's rings, are dynamically active and photometrically variable on timescales from weeks to years, with changes in dust emission \citep{xu2018,swan2019a,guidry2024} and metal line profiles \citep{wilson2014,redfield2017,manser2019}. The rapid, irregular disruption of planetesimals \citep{vanderburg2015,guidry2021,vanderbosch2021,bhattacharjee2025} and exoplanet transits \citep{vanderburg2020} have been detected via time-series photometry of white dwarfs.

\begin{figure}[t]
    \centering
    \includegraphics[width=0.9\linewidth]{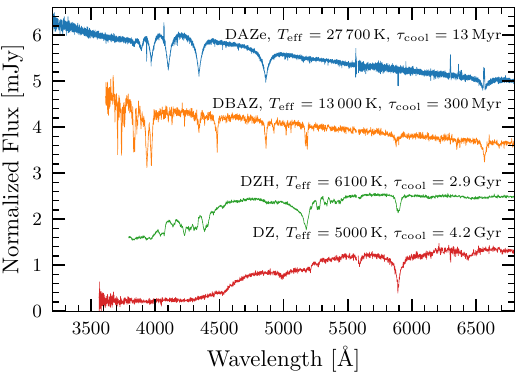}
    \caption{Four examples of metal detections in optical-blue spectra of white dwarfs. From top to bottom: H-atmosphere accreting a gaseous planet \citep[emission lines from volatile elements are visible;][]{gaensicke2019}; He-atmosphere accreting hydrated debris \citep{raddi2015}; two He-atmospheres (the top one is magnetic) accreting dry debris \citep{hollands2017}. Spectral types, effective temperatures, and cooling ages are noted in figure.}
    \label{fig:spectra}
\end{figure}

\begin{figure}[t]
    \centering
    \includegraphics[width=0.9\linewidth]{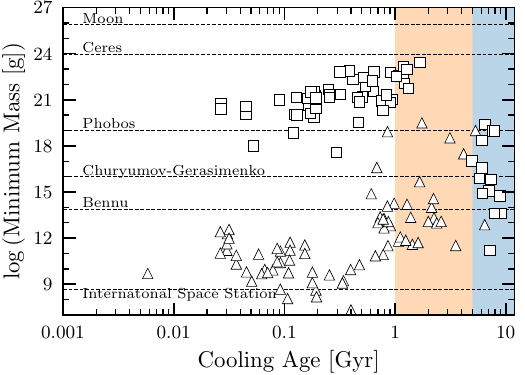}
    \caption{Minimum accreted masses inferred from the five most commonly detected elements (O, Mg, Si, Ca, Fe) as a function of white-dwarf cooling ages (i.e., the time since the white dwarf formed), for a subsample in \citet{williams2024}. Triangles and squares label white dwarfs with H- and He-dominated atmospheres, respectively. Dashed lines compare with the masses of Solar-system objects. The colored bands roughly define the cooling-age boundaries for a 0.6\,M$_\odot$ white dwarf at a distance of 100\,pc when it possesses an apparent magnitude of {\em Gaia}~$G = 18$, 20, and 23\,mag; its $\sim 2$\,M$_\odot$ progenitor would have a $\sim 2$\,Gyr pre-white dwarf age. The considered magnitude limits could be respectively reached by existing instruments on the 8-m Very Large Telescope (VLT), like UVES ($R =  20\,000)$, X-shooter ($R = 10\,000)$, and FORS2 ($R = 1000$), delivering a signal-to-noise ratio of $\mathrm{S/N} \approx 10$/pixel in optimal observing conditions with 2-hr exposures.}
    \label{fig:masses}
\end{figure}

Most data on white-dwarf planetary systems come from the analysis of relatively low-resolution spectra (Fig.\,\ref{fig:spectra}), like those obtained by the Sloan Digital Sky Survey \citep[][]{koester2011,hollands2017}. The accretion of planetary debris (aka \textit{``metal-pollution''}) enriches with metals the otherwise pure H or He atmospheres of white dwarfs, producing absorption lines from rock-forming elements whose abundances resemble the variety of Solar-system asteroids \citep{gaensicke2012}. These signatures persist from days to Myr depending on atmospheric diffusion timescales set by temperature, density, composition, and convection-zone depth \citep[][]{koester2009,cunningham2021}. Fig.\,\ref{fig:masses} illustrates the minimum accreted mass of planetesimals that is inferred from the abundances of detected elements. 

Most white dwarf atmospheres accrete rocky, volatile-poor, differentiated debris \citep{swan2019b}, but excess O and H indicate the disruption of hydrated bodies \citep{farihi2013,raddi2015,gentilefusillo2017,trierweiler2025}.  More rarely, volatile-rich debris accreted from Kuiper-belt-like bodies  \citep{xu2017}  or giant-planet atmospheres \citep{gaensicke2019} have been observed. Recent detections of Li \citep{hollands2021,kaiser2021} and Be \citep{klein2021} may even provide the first evidence for the most ancient planetesimals in our Galaxy or those formed in harsh environments.

The analysis of white dwarfs in wide binaries suggests that accreted planetary debris trace the primordial chemistry of their non-degenerate stellar companions, enabling studies of planet formation in diverse Galactic populations 
\citep[][]{bonsor2021}, but also that chemical anomalies or heavy enrichment may imply different formation and evolutionary pathways \citep{noor2024,aguilera2025}. 
\vspace{-0.1cm}
\subsection{Occurrence of evolved planetary systems}

Atmospheric enrichment from planetary debris is inferred for about 25–50\% of white dwarfs cooler than $27\,000$\,K \citep{zuckerman2003,koester2014,manser2024} corresponding to a few 10\,Myr since the white dwarf's birth. Most detections typically show 1--2 elements,  but only a few hundred stars exceed four elements due to spectral-coverage, resolution, and data quality \citep{williams2024}. Dusty discs occur in 1--4\% of the metal-enriched white dwarfs \citep{rocchetto2015,rebassa2019,wilson2019}. On the other hand, the gaseous discs are observed for 1--4\% of those showing a dusty component \citep{manser2020}. About 1\% of the actively accreting white dwarfs monitored by the Zwicky Transit Facility (ZTF) also show transits from disrupted planetesimals \citep{robert2024}. A few Jupiter-sized planets have also been found, including recent {\it JWST} detections \citep{limbach2024,mullally2024}. 
\vspace{-0.1cm}
\section{Key Science Drivers in the 2040s}
\label{sec:openquestions}
The central goals for the study of evolved planetary systems at white dwarfs are a robust, unbiased census of their occurrence, chemical diversity, architecture, and variability. 

The ESA {\em Gaia} mission currently provides the first, statistically significant sample of white dwarfs down to $G \approx 20$\,mag \citep[$\approx 360\,000$ highly reliable candidates;][]{gentilefusillo2021}. Thus, scaling the inferred occurrence of evolved planetary systems, we expect that up to $\sim$100\,000 will be disguised among the {\em Gaia} white dwarf population. A few thousands will be uncovered in the next 5--10 years by multi-object spectroscopic surveys on 4-m class telescopes, e.g., DESI \citep{cooper2023}, 4MOST \citep{dejong2019}, WEAVE \citep{jin2024}, and SDSS-V \citep{kollmeier2025}. However, those systems at the faint end of the magnitude distribution ($G \gtrsim 19$\,mag) and the weakly accreting objects may still remain undetected. 

Existing ground-based facilities available to the ESO community in the southern hemisphere, such as UVES and X-shooter, will allow follow-up of the most interesting individual objects, granting low-to-high resolution and broad optical/near-infrared coverage. The {\it Hubble Space Telescope}, which has been fundamental to deliver ultraviolet spectra for a few hundred of the hottest (more recently formed) metal-enriched white  \citep[e.g.,][]{ouldrouis2024} in the past 30 years, is expected to survive until the 2030s; however, its ultraviolet instruments have significantly degraded making their future use uncertain. Also from space, at near- and mid-infrared wavelengths, {\em JWST} is contributing to the study of those systems possessing circumstellar discs \citep{swan2024,farihi2025} or giant planets. Looking ahead, {\em JWST} could in principle be employed to search for molecular biosignatures in the atmospheres of white-dwarf exoplanets \citep{kaltenegger2020}.

In the coming decade, the Legacy Survey for Space and Time (LSST) at the Vera C. Rubin Observatory \citep{zekljko2020} is predicted to identify up to a few  million white dwarfs down to $r = 23$--25\,mag via combinations of accurate parallaxes, proper motions, and deep-stacked images in six optical bands  \citep[$ugrizy$;][]{fantin2020}. Such large numbers of newly discovered white dwarfs -- potentially reaching up to 3\,kpc away from the Sun  -- will allow the study of evolved planetary systems in the distinct Galactic populations (thin/thick disk and halo). Moreover, based on the ZTF observations, a few thousand disrupting minor-bodies and exoplanets could be detected from the analysis of the fast, time-domain photometry of LSST.

In addition to the existing instruments, a few new ground-based or space-borne facilities will be suitable for the study white-dwarf planetary systems in the 2030s and beyond: 

{\bf I.}  The {\it Ultraviolet Explorer} \citep[{\it UVEX};][]{kulkarni2021}, to be launched in the 2030s, will observe the full sky in the near- and far-UV, improving the spatial resolution and spectroscopic sensitivity of its predecessor {\em GALEX} \citep{morrissey2007} and reaching the 26$^\mathrm{th}$ magnitude in the near-UV; 

{\bf II.} Also in the 2030s, the NASA's concept mission {\em PRobe far-Infrared Mission for Astrophysics} \citep[{\em PRIMA};][]{moullet2023} could enable to identify water vapor emission in roughly 230 metal-enriched white dwarfs within 60 pc away from the Sun \citep{okuya2025}.
   
{\bf III.} The Cassegrain U-Band Efficient spectrograph \citep[CUBES;][]{cristiani2022}, scheduled for first light at the VLT in the 2030s, will enable the detection of metal transitions in the 300--400\,nm range reaching a S/N of 20 in 1-hr for targets of 18--20\,mag with resolving powers of $R = 20\,000$ and 7000, respectively; 

{\bf IV.} Also in the 2030s, the ANDES spectrograph \citep{maiolino2013} of the Extremely Large Telescope will permit  very high resolution ($R \sim 100\,000$) observations of the faintest, coolest white dwarf targets, i.e., potentially the oldest planetary hosts in our Galaxy; 

{\bf V.} In the 2040s, the NASA's concept mission {\em Habitable Worlds Observatory} ({\em HWO}) will target a few hundred bright white dwarfs at high spectral resolution in the far-UV, searching for volatile elements in those white dwarfs that show evidence of planetary debris accretion \citep{xu2025}.
\vspace{-0.1cm}
\subsection{Expanding the census of evolved planetary systems}
While the technological landscape in the 2030s is well defined, ESO's call for Transforming Astronomy in the 2040s brings in an opportunity for the community to propose new facilities and build synergies with other future instruments.  

Aiming to achieve an unbiased census of the occurrence of white-dwarf planetary systems and to map their chemical diversity, architecture, and variability, we identify a number of requirements for a suitable spectroscopic facility: {\bf (i)} broad, blue to near-infrared coverage (ideally between 3000--10\,000\,\AA); {\bf (ii)} high sensitivity at blue wavelengths ($3000$--$4000$\,\AA) to cover some strong metal lines, but also the rarest accreted metals; {\bf (iii)} multi-resolution capability ($R = 5000$–$20\,000$); {\bf (iv)} massive multiplexing for population studies; {\bf (v)} time-domain reactivity and the possibility to observe variability over timescales from hours to years. 

High-sensitivity in the blue and broad wavelength coverage are characteristics of planned or existing single-target instruments, but the multi-object survey instruments on 4-m class telescopes do not have both these capabilities. Similarly, the timely, repeated follow-up of targets over long timescales is not always feasible on single target facilities. Thus, a new facility combining some -- if not all -- the required characteristics would fill a critical gap in current technology for the first time. It will not only uncover vast numbers of metal-enriched white dwarfs via low-resolution, follow-up spectra of LSST candidates (down to $G = 23$--25\,mag), but also precisely characterize their accreted material and monitor the variability via time-domain, high-resolution spectroscopy for the brightest ones ($G \leq 20$\,mag). A rapid-response capability is also crucial to track the real-time disruption of planetesimals and to map the kinematics of circumstellar gas via changes in metal absorption and emission lines. Such a future facility should enable \textit{``industrial scale''}, prompt follow-up, providing timely observations of transient planetesimal disruptions and delivering the demographic census that is needed to deeply understand the post-main sequence evolution of exoplanets. 
\vspace{-0.1cm}
\subsection{Technology and Data Handling Requirements}
\label{sec:tech}
Absorption lines from 24 different metals have been detected in the atmospheres of white dwarfs enriched by planetary debris, but mostly 1--4 elements are typically observed in addition to the hydrogen's Balmer series and the He~{\sc i} lines \citep[][]{williams2024,xu2025}. The intensity and appearance of metal lines vary with the bulk of accreted material, effective temperature, and dominant element in the white dwarf atmosphere (Fig.\,\ref{fig:spectra}), and their detectability strongly depends on spectral resolution and S/N. At optical wavelengths, the strongest lines can be the Mg\,{\sc i} triplets (near 3829 and 5172\,\AA), the Ca\,{\sc ii} H\&K lines (3933/3969\,\AA), and the Na\,{\sc i}-D doublet (5890/5896\,\AA); elusive elements (e.g., Be, Ti, V, Mn, Ni, Cu) may be detected in the 3000--3600\,\AA\ spectral region. In the near-infrared, instead, the O\,{\sc i} triplet (7773\,\AA) is key for measuring the accretion of hydrated rocks, while the Ca\,{\sc ii} triplet (8498, 8542, and 8662\,\AA) is a strong tracer of gas accretion when it is seen in emission. 

In order to identify new white dwarfs  displaying traces of planetary debris accretion, among the new objects identified by LSST, the proposed facility should be capable to observe many objects at the same time (e.g., being equipped with a multi-object spectrograph that could work in survey mode) or allow the follow-up of many objects via an array of telescopes. It should offer a low-resolution mode ($R \sim 5000$) that enables the identification and characterization of metal enriched white dwarfs among the larger available samples in the 2040s. The low-resolution mode should have a broad wavelength coverage, ideally ranging from the atmospheric cutoff to the near-infrared (3000--10\,000\,\AA). Higher-resolution ($R \gtrsim 20\,000$) will be key to detect minute traces of heavy elements, in order to determine robust statistics on the occurrence of exoplanetary systems at white dwarfs. Spectral windows centered on the most frequently detected, strongest lines in the optical and near-infrared would be suitable (e.g., the Ca\,{\sc ii} lines).

Both low- and high-resolution capabilities should be enhanced by a rapid response mode (e.g., via ToO programmes), enabling to observe the growing numbers of white dwarfs that display transits from disrupted asteroids and planetesimals. This will allow follow-up of the most interesting systems via time-resolved spectroscopy, monitoring their variable debris accretion on short and long timescales. 

Although the number of newly discovered white dwarfs will increase by an order of magnitude and time-resolved observations will be needed, the expected data loads are much smaller compared to those predicted for other Galactic and extragalactic targets. Therefore, the necessary tools for handling them will be already available in the 2040s.

\begin{acknowledgements}
 RR acknowledges support from Grant RYC2021-030837-I and MEC acknowledges grant RYC2021-032721-I, both funded by MCIN/AEI/ 10.13039/501100011033 and by “European Union NextGeneration EU/PRTR”. This research was partially supported by the AGAUR/Generalitat de Catalunya grant SGR-386/2021 and the Spanish MINECO grant, PID2023-148661NB-I00. 
\end{acknowledgements}


\bibliographystyle{aa}

\bibliography{references}

\end{document}